\documentstyle[titlepage,12pt]{article}
\def\half{\frac{1}{2}}

\def\sgn{\mbox{sgn}}
\def\be{\begin{equation}}
\def\grad{\nabla}
\def\ee{\end{equation}}
\def\rq#1{(\ref{eq#1})}
\def\Tac{{\cal T}}
\newcount\sectionnumber
\def\sect
{\global\equationnumber=0
\global\advance\sectionnumber by 1
\the\sectionnumber . }
\newcount\equationnumber
\def   \num
{\eqno{\global\advance\equationnumber by 1
\left(\the\sectionnumber .\the\equationnumber \right)}}
\begin{document}

\title{Properties of Asymptotically Flat Two-Dimensional Black Holes}
\author{R.B. Mann, M.S. Morris and S.F. Ross\\
        Department of Physics\\
        University of Waterloo\\
        Waterloo, Ontario \\
        N2L 3G1}

\date{October 15, 1991\\
WATPHYS TH-91/04}
\maketitle

\begin{abstract}
We investigate properties of two-dimensional asymptotically flat black
holes  which arise in both string theory and in scale invariant theories of
gravity.  By introducing matter sources in the field equations we show
how such objects can arise as the endpoint of gravitational collapse. We
examine the motion of test particles outside the horizons, and show that
they fall through in a finite amount of proper time and an infinite amount
of coordinate time. We also investigate the thermodynamic and quantum
properties, which give rise to a fundamental length scale. The 't Hooft
prescription for cutting off eigenmodes of particle wave functions is shown
to be source dependent, unlike the four-dimensional case.  The relationship
between these black holes and those considered previously in $(1+1)$
dimensions is discussed.
\end{abstract}

\section{Introduction}

The study of black hole spacetimes in two-dimensional theories of gravity
\cite{BHT,MST,MFound} has proven to be a useful exercise in exploring a variety
of
issues associated with both classical \cite{Arnold} and quantum gravity
\cite{semi,SharTomRobb}. Recently there has been some interest in two
dimensional
spacetimes described by the metric
\be
ds^2 = -(1-a e^{-Qx})dt^2 + \frac{dx^2}{1-a e^{-Qx}} \quad . \label{eq1}
\ee
where $a$ and $Q$ are constants of integration.
Although first discovered as a solution to a scale-invariant
higher-derivative theory of gravity \cite{JurgJMP}, this metric has been
found to be a solution to $c=1$ Liouville gravity
\cite{L+R} as well as to a non-critical string theory in two spacetime
dimensions \cite{MSW,EDW}. In this latter case, the
theory may be considered to be a critical string theory in a non-trivial
background; setting to zero the one-loop beta function of the bosonic
sigma model with two target spacetime dimensions yields field equations for
the metric and dilaton fields whose solution is \rq1 and
\be
\Phi = -\frac{Q}{2}x    \label{eq2}
\ee
where now $Q^2=8/\alpha'$, $\alpha'$ being the string coupling constant.
The effective (Lorentzian) target space action from which this solution
follows is
\be
S=\int d^2xe^{-2\Phi}\sqrt{-g}(R - 4(\nabla\Phi)^2 + c), \label{eq2a}
\ee
Alternatively, this solution may also be  described by a $k=9/4$ gauged
SO(2,1)/SO(1,1) WZW model \cite{EDW}.

The metric \rq1 resembles the $(3+1)$ dimensional Schwarzschild black hole
in a number of ways. It is asymptotically flat as $x\to\infty$ and it
has  an event horizon at $x=\ln(a)/Q$ (provided $a>0$). The curvature
scalar is
\be
R = -aQ^2 e^{-Qx}  \label{eq3}
\ee
which is singular as $x\to -\infty$.

For $Q<0$ analogous properties hold under inversion of the spatial
coordinates ({\it i.e.} $x$ replaced by $-x$). In this respect it is quite
unlike the $(3+1)$ dimensional Schwarzschild black hole (and the black hole
spacetimes studied previously in $(1+1)$ dimensions \cite{BHT,MST}) as the
solution is not symmetric about the origin. Such a property makes it
difficult to understand the solution \rq1 arising as the endpoint of
gravitational collapse of localized $1+1$ dimensional matter \cite{Arnold}.
Indeed, collapse to a metric of the form \rq1 would have to be ``one
sided'': all the matter would need to collapse in the same direction.
Rather one expects a general localized distribution of matter to collapse
inwardly toward some fixed point (which may be taken to be the origin),
leaving the same vacuum spacetime for {\it both} $+x$ and $-x$.

Hence in a manner analogous to the situation in $3+1$ dimensions one
expects
\be
ds^2 = -\alpha(|x|)dt^2 + \frac{dx^2}{\alpha(|x|)}  \label{eq3a}
\ee
to be the form of the exterior metric \cite{Arnold}, inversion symmetry in
the coordinate $x$ being the $1+1$ dimensional analogue of spherical
symmetry. Of course the coordinate $x$ in (\ref{eq1}) could well be
replaced by a coordinate $r$ such that $(1 - a e^{ -Qx}) = \tanh^2 r $;
then the metric (\ref{eq1}) becomes manifestly symmetric under inversion of
the new coordinate $r$. Although the Penrose conformal diagram for this
spacetime is visually identical with that for the Schwarzschild spacetime
(there being a symmetry about the vertical in the Penrose diagram) this
symmetry is {\it not} the spherical symmetry which would correspond to that
arising from spherically-symmetric gravitational collapse. Rather it is a
symmetry in the $(u,v)$ null coordinates which is present as a consequence
of the Kruskal-Szekeres extension of (\ref{eq1}) (and, in $3+1$ dimensions,
of the extension of the usual Schwarzchild metric). Note that
in the Schwarzschild
case, every point on the Penrose diagram corresponds to a 2-sphere, whereas
in the metric \rq1 every point on the Penrose diagram
corresponds to a single point. The above arguments suggest that
a better analog to Schwarzschild,
based on our work on gravitational collapse, would be for every point on
the Penrose diagram to correspond to a 0-sphere (that is, two points).
If every point corresponds to a 0-sphere the metric is symmetric in the
spatial coordinate and the solution could plausibly arrive from
gravitational collapse of localized matter.

In the bosonic sigma model
which gives rise to \rq1, the only ``matter'' field present is the tachyon.
However it is possible to construct more  complicated models in which many
other matter fields are present.  In such  models one might expect a
breakdown of conformal invariance, generating  an effective stress-energy
tensor for the $1+1$ dimensional  gravitational field.

In this paper we consider the action (\ref{eq2a}) (and its extension to
include tachyons) from a relativist's viewpoint, by incorporating effects
due to matter sources represented by a stress-energy tensor $T_{\mu\nu}$.
In the absence of tachyons, we show that the effective action mentioned
above is equivalent to that of a massive scalar field non-minimally coupled
to curvature. Apart from its relationship to string theory, we regard this
theory (coupled to matter) as being of intrinsic interest as a two-
dimensional theory of gravitation \cite{BanksL}. We also investigate the
circumstances under which a black hole in this theory might form as the
endpoint of gravitational collapse of some (finite) distribution of matter.
Specifically, we consider $1+1$ dimensional dust coupled to the
graviton-dilaton system and show that it collapses to the exterior black hole
solution
\be
ds^2 = -(1-a e^{-Q|x|})dt^2 + \frac{dx^2}{1-a e^{-Q|x|}} \label{eq4}
\ee
with the dilaton field
\be
\Phi = -\frac{Q}{2}|x|    \label{eq5}
\ee
provided appropriate surface stresses and dilaton surface charges are
included, in contrast to $(1+1)$-dimensional dust collapse to a
black hole previously studied in another context \cite{Arnold}
where no such stresses need be introduced. Apart from this, we
find that other properties are similar to the situation studied in
ref. \cite{Arnold}:
the formation of the event horizon takes a finite amount of
proper time and an infinite amount of co-ordinate time, and the interior
solution to which the solution (\ref{eq4},\ref{eq5}) is matched collapses
to a point within finite proper time.

Inclusion of a non-zero tachyon field implies that the metric and  dilaton
must be
\cite{VerDijk}
\be
ds^2 = \half(k-2) [dr^2-4 (\coth^2\frac{r}{2}-
\frac{2}{k})^{-1}d\tau^2] \label{veq1}
\ee
\be
\Phi = \log[\sinh r (\coth^2 \frac{r}{2} -\frac{2}{k})^{1/2}] \label{veq2}
\ee
provided the $L_0$ operator is equated with the target space Laplacian.
These reduce to the metric and dilaton (\ref{eq1},\ref{eq2}) in the limit
$k\to\infty$. A coordinate transformation will put this metric in the form
\be
ds^2 = -\left(1-\frac{k}{1+\cosh(Q x)}\right) dt^2 + \frac{dx^2}
{\left(1-\frac{k}{1+\cosh (Q x)}\right)}, \label{veq30}
\ee
which is symmetric in $x$. This metric and dilaton do not necessarily represent
an exact solution of the full field equations with the tachyon $\Tac\neq 0$.
However, the fact that they are invariant under inversion of $x$ makes them
worthy of study from our point of view.

Finally, we
investigate the properties of the black hole spacetimes \rq4 which arise in
such a theory, comparing them to those $(1+1)$ dimensional spacetimes
studied previously \cite{MST,Arnold}. We demonstrate that freely falling
test particles fall through the event horizons in finite proper time and
infinite coordinate time (as in the previously studied case \cite{MST}),
although the forms of the exact solutions of the equations of motion
differ.   The Hawking temperature and entropy of the horizons are computed
and finally the quantum properties are studied in the context of the 't
Hooft ``brick wall'' model. The thermodynamic properties are observed to be
identical to that of the black hole studied in \cite{MST}, and we show that
the ``brick wall'' prescription for cutting off the wavemodes of particle
wave functions near the horizon leads to a source-dependent cutoff
distance, whereas in four dimensions the cutoff distance is
source-independent.  We also study the properties of the metric (\ref{veq30})
obtained from considering tachyons, with similar results. We summarize our
results and discuss further areas of interest in a concluding section.

\section{Two-Dimensional Gravitational Actions}

In \cite{JurgJMP}, a solution is developed which has the characteristics of a
black hole. The proposed Lagrangian is ${\cal L}=R^{\kappa+1}$, for which
the field equation is
\be
R^{\kappa+1}+(1+\frac{1}{\kappa})\nabla^2 R^\kappa = 0.  \label{neq1}
\ee
The black hole is obtained by taking the limit $\kappa\to 0$. If the metric is
written in the form
\be
ds^2 = -\alpha dt^2 + \alpha^{-1} dx^2,  \label{eq6}
\ee
then the solution is given by \rq1, so that
\be
\alpha(x) = 1- ae^{-Qx}.    \label{eq7}
\ee
where $a$ and $Q$ are arbitrary constants.

This solution may also be obtained from a string theory action \cite{MSW,EDW}.
With the tachyon field taken to be $\Tac=0$, setting to zero the one-loop beta
functions of the associated sigma model effective action yields the field
equations
\be
R_{\mu\nu} - 2\nabla_\mu\nabla_\nu\Phi = 0, \label{eq8}
\ee
\be
R+4(\nabla\Phi)^2 - 4\nabla^2\Phi + c = 0, \label{eq9}
\ee
where $a$ is still arbitrary, but $c=Q^2=8/\alpha'$. In a gauge where the
dilaton is $\Phi = -Qx/2$, \rq7 is a solution of this system.

We pause here to comment that the action (\ref{eq2a}) may be recast
in the form of a Klein-Gordon action with a non-minimal coupling to
curvature \cite{semi,BanksL}. If we write $\psi=e^{-\Phi}$, then the target
space action (\ref{eq2a}) becomes
\be
S = \int d^2x\sqrt{-g}(\psi^2 R - 4(\nabla\psi)^2+c\psi^2), \label{eq11}
\ee
and variation with respect to $\psi$ yields the equation
\be
(\alpha\psi')'-\alpha^{-1}\ddot{\psi}-\frac{c}{4}\psi - \frac{1}{4}\psi R = 0.
\label{eq12}
\ee
But \rq{12} is equivalent to the (non-minimal) Klein-Gordon equation,
\be
(\alpha\phi')'-\alpha^{-1}\ddot{\phi} - m^2\phi + \xi\phi R = 0, \label{eq13}
\ee
if we identify $c/4=m^2$ and let the arbitrary coupling $\xi=-1/4$. Thus,
this system may be interpreted as a massive scalar field $\psi = e^{Qx/2}$
non-minimally coupled to curvature. We concentrate on this viewpoint, as it
permits a relatively unambiguous interpretation of the gravitational
effects, and it is easy to introduce a suitable matter term  $8\pi G
g^{\mu\nu}T_{\mu\nu}$ into the target space action. Such a term models the
breakdown of conformal invariance in the context of the string action
\rq{10} .

The effective (Lorentzian) target space action
(\ref{eq2a}) becomes
\be
S_{eff} = \int d^2x \sqrt{g}\left[e^{-2\Phi}\left(R - 4(\grad\Phi)^2 +
   (\grad\Tac)^2  + V(\Tac) +c \right) \right]
   \label{eq10}
\ee
upon inclusion of the tachyon field. The equations which follow from
the action \rq{10} have the solution (\ref{veq30}), where now
\be
\alpha(x)=\left(1-\frac{k}{1+\cosh (Q x)}\right) \quad  \label{eq10a}
\ee
for an appropriate choice of $V(\Tac)$, although the metric and dilaton
fields (\ref{veq1},\ref{veq2}) do not necessarily represent an exact
solution of the full field equations \cite{VerDijk}.

We wish to compare the black hole solutions considered in this paper to
those found in the minimally coupled theory previously developed in
\cite{MST,Arnold}. In this simpler theory, the field equation is
\be
R=8\pi GT,     \label{eq14}
\ee
where $R$ is the Ricci scalar and $T$ is the trace
of the stress-energy tensor.  This field equation follows from the action
\cite{semi}
\begin{equation}
S=\int d^2x\,\sqrt{-g}g^{\mu\nu} \left(\frac{1}{2}
\partial_\mu\Psi\partial_\nu\Psi + \Psi R_{\mu\nu}-
8 \pi G T_{\mu\nu} \right)\label{eq15}
\end{equation}
where $\Psi$ is an auxiliary field, whose equation guarantees the
conservation of the stress-energy tensor $T_{\mu\nu}$.
If we write the metric in the form \rq6, and take the stress-
energy to be that of a point mass at the origin,
then there is a black hole solution to this
equation,
\be
\alpha(x)=2M|x|-1, \label{eq16}
\ee
with horizons at $|x|=1/(2M)$.  Comparison with the Newtonian limit
\cite{MFound} indicates
that the parameter $M$ may be interpreted as the
mass of the black hole. This solution is examined in more detail in
\cite{MST}.

\section{Black Holes and Gravitational Collapse}

As mentioned in the introduction, the  solution \rq7 has been referred to as
a black hole, as it is asymptotically flat and has an event horizon where
$\alpha(x)=0$, at $x=\ln(a)/Q$ (provided $a>0$).
As we shall see, a combination of the parameters $a$ and $Q$ will
determine the mass of the black hole.
The solution also
has a curvature singularity  $R \to -\infty$ as $x \to -\infty$ (assuming
$Q>0$). It is worth noting at this point that the sign convention for $R$
implicit in the equations (\ref{eq8},\ref{eq9}) is the opposite of that
used in \cite{MST}.

However the asymmetric black hole \rq7 differs from the
$(3+1)$-di\-men\-sional  Sch\-warz\-child case
in several ways, which make it difficult to interpret.  It is asymmetric in
the spatial variable, has only one event horizon, and has a curvature
singularity as $x\to -\infty$ (as opposed to $r\to 0$). One might expect
that such differences would be absent for a black hole which arises from
the collapse of a distribution of matter coupled to  the theory described
by \rq{11}.

Introducing a source of stress-energy into the action \rq{11},
the field equations are
\be
e^{-2\Phi}(R_{\mu\nu} + 2 \grad_\mu \grad_\nu \Phi) = 8\pi T_{\mu\nu}
\label{eq17}
\ee
\be
R - 4(\grad\Phi)^2 + 4 \grad^2\Phi + c = 0   \label{eq18}
\ee
where we have written $\psi=e^{-\Phi}$.  In the context of the string
action \rq{10}, $c= (26-D)/(3\alpha') = 8/\alpha'$ since $D=2$, and the
tachyon has been set to zero. It is straightfoward to show that the system
(\ref{eq17},\ref{eq18}) guarantees the covariant conservation of
$T^{\mu\nu}$.

Consider a symmetric distribution of dust centered about the origin over a
finite spatial range.   The solutions to the field equations
(\ref{eq17},\ref{eq18}) in the absence of matter are
\be
ds^2 = -(1-a e^{-Q|x|})dt^2 + \frac{dx^2}{1-a e^{-Q|x|}} \label{eq19}
\ee
and
\be
\Phi = -\frac{Q}{2}|x|     \label{eq19a}
\ee
with $Q^2 = c$. This solution is valid everywhere except at $x=0$.

We next consider the interior solution, here taking the metric to
be of the form
\be
ds^2 = -d\tau^2 + {\cal R}^2(\tau) dy^2 \label{eq20}
\ee
with the stress-energy tensor
\be
T^{\mu\nu} = \rho(\tau) u^\mu u^\nu  \quad . \label{eq21}
\ee
Using comoving coordinates, the conservation laws imply that $\rho {\cal R}$
is a constant, which may be chosen to be $\rho_0$, the initial density
of the dust at $\tau=0$.

With this choice of coordinates the field equations \rq{17} become
\be
\ddot{{\cal R}} - 2 \dot{{\cal R}}\dot{\Phi} = 0  \label{eq22}
\ee
\be
e^{-2\Phi}\left(-\frac{\ddot{{\cal R}}}{{\cal R}} + 2\ddot{\Phi}\right) = 8\pi
G \rho
\label{eq23}
\ee
whereas \rq{18} becomes
\be
2\frac{\ddot{{\cal R}}}{{\cal R}} +4\dot{\Phi}^2 - 4\frac{\dot{{\cal R}}}{{\cal
R}}\dot{\Phi}
-4\ddot{\Phi} + Q^2 = 0  \quad . \label{eq24}
\ee
Insertion of \rq{22} into \rq{24} yields
\be
\Phi(\tau) = \Phi_0 - \ln\left(\cos(\frac{Q}{2}\tau)\right) \label{eq25}
\ee
which, upon substitution into \rq{23} implies
\be
{\cal R}(\tau) = 1 -\lambda \tan(\frac{Q}{2}\tau)  \label{eq26}
\ee
where $\lambda$ is a constant of integration. Eq. \rq{23} also implies that
\be
\rho_0 = \frac{Q^2}{16\pi G} e^{-2\Phi_0}   \label{eq27}
\ee
as a consequence of the conservation law.

The interior solution given by (\ref{eq25},\ref{eq26}) exhibits collapse
in a finite amount of comoving coordinate time $\tau_c =
\frac{2}{Q} \tan^{-1}(\frac{1}{\lambda})$. Since the density of dust
varies inversely with ${\cal R}(\tau)$ we see that, as in $3+1$ dimensions,
the dust collapses to a state of infinite density at this time. The
curvature
\be
R = - \frac{Q^2\lambda \sec^2(\frac{Q}{2}\tau)
         \tan^2(\frac{Q}{2}\tau)}{1 -\lambda \tan(\frac{Q}{2}\tau)}
                      \label{eq28}
\ee
also diverges at $\tau_c$.

That the gravitational collapse of dust is observable from outside may be
shown by matching the metric and the dilaton field at the edge of the dust
$x=s(\tau)$.  These conditions  are easily met by choosing
\be
|s(\tau)| = x_0 + \frac{2}{Q}\ln\left(\cos(\frac{Q}{2}\tau)\right)
\label{eq35}
\ee
which guarantees the dilaton matching, and
\be
\dot{t}^2 (1-ae^{-Q|s|}) -\frac{\dot{s}^2}{(1-ae^{-Q|s|})} = 1 \label{eq36}
\ee
which ensures that the metrics (\ref{eq19},\ref{eq20}) match at the
boundary of the dust. This latter equation has the solution
\be
t = \frac{2}{Q}\tanh^{-1}\left[
\frac{\sin(\frac{Q}{2}\tau)}{\sin(\frac{Q}{2}\tau_H)}\right] \label{eq37}
\ee
where $\tau_H = \frac{2}{Q}\cos^{-1}(\sqrt{a} e^{-Q x_0/2})$ is the comoving
time at which the black hole event horizon forms. Note that
this appears to take place at $t=\infty$ to an external observer.

It is clear that (\ref{eq35}) guarantees a $C^0$-matching of the dilaton
field inside and outside of the dust, in general suggesting the presence of
a surface stress-energy tensor.
The discontinuity in the normal derivative of $\Phi$ should also
be indicative of a surface dilaton charge, with a self-repulsion that is
balanced by a mechanical force.  In $N>1$ spatial dimensions such self-
repulsion can be balanced by surface tension; for example when $N=3$ a
conducting balloon ({\it i.e.} a 2-surface) with electrostatic charge
respects $\pi r^2 \sigma \sim \hat{T} r$ where $\sigma$ is the electrostatic
repulsive force per unit area and $\hat{T}$ is the surface tension.
However in one dimension the boundary of any object is 0-dimensional: hence
there is no analog of $\hat{T}$, necessitating a pressure normal to the edge
which balances the self-repulsion due to the dilaton surface charge.

These may be computed
by integrating the field equations (\ref{eq17},\ref{eq18}) across the dust
edge as follows.  Denoting the 2-velocity of the edge and its normal by
$u^\mu$ and $n^\mu$ respectively yields from \rq{17}
\begin{eqnarray}
\left[K\right] &=& -8\pi G S_{00}   \nonumber \\
2\left[u\cdot\nabla\Phi\right] &=& 8\pi G S_{01} \label{eq37a} \\
\left[K\right] + 2\left[n\cdot\nabla\Phi\right] &=& 8\pi G S_{11} \nonumber
\end{eqnarray}
where $e^\mu_0\equiv u^\mu$ and $e^\mu_1\equiv n^\mu$. Here
\be
\left[A\right]\equiv \lim_{\epsilon\to 0}\int_{s-\epsilon}^{s+\epsilon}dn
\left(n\cdot\nabla A\right)  \label{eq37ab}
\ee
is the discontinuity in a quantity $A$ across the edge,
and
\be
S_{ij}\equiv \lim_{\epsilon\to 0}\int_{s-\epsilon}^{s+\epsilon} dn
e^{2\Phi}e^\mu_i e^\nu_j T_{\mu\nu} \label{eq37ac}
\ee
is the surface stress tensor,
evaluated by integrating the stress-energy tensor across the edge. The
extrinsic curvature $K$ is related to the Ricci scalar via the equation
\be
R = 2\left(n\cdot\nabla K - K^2\right)  \label{eq37b}
\ee
and its discontinuity $\left[K\right]$ may be directly evaluated from
(\ref{eq35},\ref{eq37}) using $u^\mu = (\dot{t},\dot{s})$ and
\be
K \equiv u^\mu u^\nu \nabla_\mu n_\nu   \quad  .  \label{eq37c}
\ee
Similarly, the discontinuities in $u\cdot\nabla\Phi$ and $n\cdot\nabla\Phi$
may be computed using \rq{35} and the exact solution
(\ref{eq25},\ref{eq26}). The results are
\be
\left[K\right] = \frac{Q}{2}\frac{1+\dot{s}^2}{\alpha({s})\dot{t}}
\left(a e^{2\Phi_0}-1\right) \label{eq37d}
\ee
\be
\left[n\cdot\nabla\Phi\right] = -\frac{Q}{2}{\alpha({s})\dot{t}}
= -\left[K\right] \label{eq37e}
\ee
upon using \rq{35}, and
\be
\left[u\cdot\nabla\Phi\right] = 0  \quad , \label{eq37f}
\ee
yielding a surface stress
\be
S_{ij} = \left(\begin{array}{cc} -\left[K\right] & 0 \\
                                    0     & 3\left[K\right]
                                    \end{array}\right)  \quad .\label{eq37g}
\ee
That a dilaton surface charge is necessary is easily seen by considering
integration of \rq{18} across the edge; in the absence of any such charge
we obtain $2\left[n\cdot\nabla\Phi\right] = -\left[K\right]$, in conflict
with \rq{37e}. The source $J$ for such a charge may be introduced by
adding a term $J(x)\psi^2(x)$ in the action \rq{11}, yielding
in place of \rq{18}
\be
R - 4(\grad\Phi)^2 + 4 \grad^2\Phi + J + c = 0   \label{eq37h}
\ee
whose matching condition is
\be
2\left[K\right] + 4\left[n\cdot\nabla\Phi\right] + 16\pi\sigma_D
= -2\left[K\right] + 16\pi \sigma_D = 0  \label{eq37i}
\ee
where $16\pi\sigma_D = \lim_{\epsilon\to 0}\int_{s-\epsilon}^{s+\epsilon} dn
J$.  Note that the presence of such a source modifies the conservation laws
so that
\be
\nabla_\nu J =  16\pi G \nabla_\mu T^\mu_\nu .\label{eq37j}
\ee
In both the vacuum and dust regions $J=0$, yielding the familiar
conservation laws.
{}From a viewpoint perhaps more palatable to string theorists, $J$
can be modelled by a
tachyonic potential $V(\Tac)$  whose form is a delta
function point source at the dust edge.

For collapse to
take place $e^{Q x_0} > a$; which from (\ref{eq25},\ref{eq27}) implies
\be
\rho_0 > \frac{a Q^2}{16\pi G}  \label{eq38}
\ee
showing that the initial density must exceed a critical value, as noted
previously for two dimensional collapse \cite{Arnold}. A light signal
emitted from the surface at time $t$ obeys the null condition
\be
\frac{dx}{dt} = (1-ae^{-Q|x|}) \label{eq39}
\ee
which arrives at a distant point $\tilde{x}$ at time
$$
\tilde{t} = t + \int_{s(\tau)}^{\tilde{x}}\frac{dt}{dx} dx
$$
\be
= t + \frac{1}{Q}\sgn(\tilde{x})\ln\left[
\frac{(e^{-Q|\tilde{x}|}/a-1)\cos^2(Q\tau_H/2)}
{\cos^2(Q\tau/2)-\cos^2(Q\tau_H/2)}\right]  \label{eq40}
\ee
which diverges as $\tau\to \tau_H$, showing that the
collapse to $R=0$ is completely unobservable for outside observers
on either side of the fluid. The corresponding redshift factor is
easily computed to be
\be
z = \frac{\cos(Q\tau/2)}{\sin(Q\tau_H/2)-\sin(Q\tau/2)} - 1 \label{eq41}
\ee
which also diverges as $\tau\to \tau_H$.

The above features of this collapse are quite analogous to those discussed
previously \cite{Arnold} for the black hole solution \rq{16} in the theory
\rq{15} .  In both cases the endpoint of this collapse may be modelled by a
point source at the origin
\be
T=(p-\rho),\,\, p=0,\,\, \rho=\frac{M}{2\pi G}\delta(x),
\label{eq29}
\ee
where $M$ is a paramter associated with the mass of the source.
In this case the exterior solution
\rq{20} for $\Phi$ becomes
\be
\Phi = \frac{M |x|}{\alpha(0)}  \label{eq30}
\ee
whereas
\be
\alpha(x) = 1 - ae^{-2M|x|}    \label{eq31}
\ee
with
\be
J(x) = 2M(a-2)\delta(x) \label{eq31a}
\ee
as the dilaton source.

Thus the metric for this black hole is
\be
ds^2 = -(1-ae^{-2M|x|})dt^2+(1-ae^{-2M|x|})^{-1}dx^2 , \label{eq32}
\ee
which is locally identical to \rq1
with $Q=2M\sgn(x)$. Horizons will only form for positive
$M$, so we will assume $M >0$ in the rest of this paper.

The metric \rq{32} may be written in a conformally flat form, by
means of the transformations
\be
\begin{array}{ccll}
y & = & \,\,\,\frac{1}{2M}\ln(e^{-2Mx}-a),& x\in(-\infty,-\ln(a)/2M),  \\
\vert y\vert& =& -\frac{1}{2M}\ln(a-e^{2M|x|}),& x\in(-\ln(a)/2M,\ln(a)/2M),\\
y & = & -\frac{1}{2M}\ln(e^{2Mx}-a),& x\in(\ln(a)/2M,\infty),
\end{array} \label{eq33}
\ee
which are only valid for $M$ positive. Each of the above intervals is
mapped to an interval $y\in(-\infty, \infty)$. The metric in the new
$(y,t)$ coordinates is
\be
\begin{array}{ccll}
ds^2 & = & (1+ae^{-2My})^{-1}(-dt^2+dy^2), & \mbox{timelike,} \\
ds^2 & = & (ae^{2M\vert y \vert} -1)^{-1}(dt^2-dy^2), & \mbox{spacelike,} \\
ds^2 & = & (1+ae^{2My})^{-1}(-dt^2+dy^2), & \mbox{timelike}
\end{array}
\label{eq34}
\ee
where ``timelike'' denotes a region of signature $(-,+)$ and ``spacelike''
denotes a region of signature $(+,-)$.
The asymptotically flat sides of the exterior patches correspond to
$x\rightarrow\pm \infty$, and in the interior patch $y(x)=0$ at $x=0$, so
the (delta-function) singularity is at the origin of the $y$ coordinate.

The black hole solution \rq{19} has an
event horizon which is a zero-sphere (a pair of
points) with the singularity in the centre of the sphere. Geodesics of
particles falling into this black hole are cut off at finite parameter
values, as we shall see in the next section.  We shall consider further
properties of this black hole metric in the next two sections, postponing
discussion of the black hole (\ref{veq30}) until section 6.

\section{Classical Properties: The Motion of Test Particles}

We now wish to consider the motion of test particles in the gravitational
field of the mass $M$. We will assume that the test particle's mass is
small compared to $M$, so that we may neglect its contribution to the
gravitational field. The geodesic equations are
\be
\frac{d^2x^\gamma}{d\lambda^2}+\Gamma^{\gamma}_{\mu\nu}\frac{dx^\mu}{d\lambda}
\frac{dx^\nu}{d\lambda} =0,  \label{eq42}
\ee
where $\lambda$ is an affine parameter along the geodesics. In the metric
\rq{32}, these equations are
\begin{eqnarray}
\ddot{x} - \half\alpha^{-1}\alpha'\dot{x}^2 + \half\alpha\alpha'\dot{t}^2 &=&
0,
\label{eq43}\\
\ddot{t} + \alpha^{-1}\alpha'\dot{x}\dot{t} &=& 0, \label{eq44}
\end{eqnarray}
with solutions
\be
\dot{t} = 1/\alpha,\, \dot{x}^2 = 1-E^2\alpha,  \label{eq45}
\ee
where $E$ is a constant of integration related to the proper time, since
\be
d\tau^2 = \alpha dt^2 - \alpha^{-1} dx^2 = E^2d\lambda^2.   \label{eq46}
\ee
Thus for massless particles $E=0$, for massive particles $E^2>0$.

It is possible to solve the equations (\ref{eq45}) explicitly for $x$ and $t$
in terms of the proper time $\tau$. In all cases, initially infalling particles
will cross the horizon in finite proper time and infinite coordinate time.
However, particles initially moving away from the hole may escape to infinity,
as the solution is asymptotically flat, unlike the black hole (\ref{eq16}).

The solutions for massless particles, with initial conditions $x(0)=x_0$,
\\$t(0)=0$ are:
\begin{eqnarray}
|x|(\lambda)&=&-\lambda+x_0,    \label{eq47}\\
t(\lambda)&=&\lambda-\frac{1}{2M}\ln(1-ae^{-2M(-\lambda+x_0)}) \nonumber\\
 &  &+\frac{1}{2M}\ln(1-ae^{-2Mx_0}),  \label{eq48}
\end{eqnarray}
where $\lambda$ is an affine parameter along the geodesics. These solutions are
only valid for $|x|> \ln(a)/2M$. The particle reaches the horizon
$|x|=x_H=\ln(a)/2M$ when the affine parameter is $\lambda=x_0-x_H$, and the
coordinate time $t\rightarrow\infty$ at this value of $\lambda$.

For massive particles, the solution is divided into three cases. For $E^2<1$,
the solution is
\begin{eqnarray}
|x|(\tau)&=&-\frac{1}{2M}\ln(\frac{1-E^2}{aE^2}(\tanh^2(\tau)-1)),
\label{eq49}\\
t(\tau)&=&\tau-\sqrt{1-E^2}\tanh^{-1}(\sqrt{1-
E^2}\tanh(\tau)),  \label{eq50}
\end{eqnarray}
where the proper time
$\tau$ has been rescaled to simplify the expression, and the initial
condition $t(0)=0$ has been applied. At the horizon, $|x|=x_H$ and
$\tau=\tanh^{-1}(1/\sqrt{1-E^2})$. The coordinate time will diverge at this
value of $\tau$. For $E^2=1$, the solution is
\begin{eqnarray}
|x|(\tau)&=&\frac{1}{2M}\ln(a\tau^2),   \label{eq51}\\
t(\tau)&=&\tau+\half\ln(\frac{\tau-1}{\tau+1}),
\label{eq52}
\end{eqnarray}
where $\tau$ is rescaled for simplicity. At the horizon
$\tau=1$, and thus the coordinate time diverges. The third case, $E^2>1$,
is only possible close enough to the horizon so that $1-E^2\alpha(x)>0$, as
otherwise the kinetic energy of the particle will be negative. The solution
for this case is
\begin{eqnarray}
|x|(\tau)&=&-\frac{1}{2M}\ln(\frac{E^2-
1}{aE^2}(\tan^2(\tau)+1)), \label{eq53}\\
t(\tau)&=&\tau+\sqrt{E^2-1}\tanh^{-1}(\sqrt{E^2-1}\tan(\tau)),  \label{eq54}
\end{eqnarray}
using the same conditions as for $E^2<1$. As the particle crosses the
horizon, $\tau=\tan^{-1}(1/\sqrt{E^2-1})$, and thus the coordinate time
$t\rightarrow\infty$. All the above solutions assume an initially infalling
particle, that is, $\dot{x}(0)/x(0)<0$, and are valid only in the exterior
region $|x|> x_H$.

The geodesic deviation equation cannot be solved explicitly for massive
particles, but it shows the expected behaviour that the separation between
massive particles increases as they fall towards the hole. The equation is
\be
a^\sigma=R_{\mu\rho\nu}^\sigma n^\rho u^\mu u^\nu,    \label{eq55}
\ee
where $n$ is the geodesic separation vector, $a$ is the acceleration and
$u$ is the tangent to the geodesic (the two-velocity of a freely-falling
particle). This equation can be simplified by assuming $n$ orthogonal to
$u$, and assuming $u$ has unit norm for massive particles. The result is
$a^\sigma=-\half\alpha''n^\sigma$ for massive and $a^\sigma=0$ for massless
particles. As $\alpha''$ is negative everywhere, this means that nearby
geodesics will separate because of the difference in force on them.
Solution of the equation for massless particles yields
\be
n^{\sigma}=C(1,\alpha) \mbox{ or } n^{\sigma}=C(1/\alpha,-1), \label{eq56}
\ee
where $C$ is a constant of integration. These values of $n$ give the
directions of the edges of the light cone. For large $x$, $\alpha=1$, and
they are $n^{\sigma}=C(1,1)$, $n^{\sigma}=C(1,-1)$, as for flat space. At
the black hole, $\alpha=-1$, and the directions are $n^{\sigma}=C(1,-1)$,
$n^{\sigma}=C(-1,-1)$, showing that the light cones `tip over', being
turned 90 degrees from the flat space directions at the singularity, in
analogy with the four-dimensional case.

\section{Quantum Properties}

We now consider the thermodynamic properties of the horizon, as in
\cite{MST}. The radiation temperature of the horizon is most easily
calculated by transforming the horizon to the origin of a polar coordinate
system \cite{BHT}, and requiring that the euclideanized time coordinate be
periodic. The metric then becomes
\be
ds^2=\alpha(x(r))d\tau^2+dr^2,   \label{eq57}
\ee
where $r$ is a coordinate such that $dx/dr=\sqrt{\alpha(x)}$,
which is well-defined as long as $\alpha\geq0$. Expanding about $r=0$ gives
\be
\alpha(r)= \alpha(0)+r\sqrt{\alpha(0)}\alpha'(0)+\half r^2\left(\alpha''(0)
\alpha(0)+\half(\alpha'(0))^2\right)+\cdots . \label{eq58}
\ee
If we choose $\alpha(0)=0$, that is, the event horizon is at the origin,
equation (\ref{eq57}) becomes
\be
ds^2=r^2\left(\frac{\alpha'(0)}{2}\right)^2 d\tau^2+dr^2,  \label{eq59}
\ee
and imposing the requirement that $\tau$ be periodic yields the temperature
\be
T=\frac{\hbar}{2\pi}\left|\frac{\alpha'(x_H)}{2}\right|.   \label{eq60}
\ee
For the black hole given by the metric \rq{32}, the temperature is
\be
T=\frac{\hbar}{2\pi}M,   \label{eq61}
\ee
which is the same as for the simple black hole given by (\ref{eq16})
\cite{MST,semi}.

The entropy may now be deduced from the thermodynamic equation proposed by
Bekenstein \cite{entropy},
\be
d {\cal M}=TdS+\Phi dQ,   \label{eq62}
\ee
where $\Phi$ plays the role of an electric potential, $Q$ (here only)
designates the electric charge, and there is no angular
momentum term in two dimensions. Identification of $ {\cal M} $,
the thermodynamic mass, leads to an interesting ambiguity. We
can evaluate $ {\cal M} $ by the quasi-local mass formula of \cite{napyost},
Eq.\ (8.12). We find
\be
{\cal M} = 2 a M e^{ - 2 \Phi_{0} } ,          \label{eq62a}
\ee
where we have reinserted for the moment a dependence on the
integration constant for the dilaton, $ \Phi_{0} $. McGuigan {\it et al}
note that this result is in agreement with earlier ADM determinations
of the mass if they set $ a = 1 $.
Now, knowing the radiative temperature, we can find the entropy by integrating
\be
T^{-1}=\frac{\partial S}{\partial {\cal M}}.   \label{eq63}
\ee
However an ambiguity arises because there is a question of interpretation
in determining the dependence of the
temperature on $ {\cal M } $. For McGuigan {\it et al} \cite{napyost},
$M$ is a constant (it is a parameter set by the string theory), and thus,
the temperature itself is a constant, independent of the thermodynamic mass
$ {\cal M} $. If we follow this lead, then we find that the black-hole
entropy is proportional to the thermodynamic mass
\be
S_{bh} = \frac{2\pi}{\hbar M} {\cal M} .        \label{eq63a}
\ee
Alternatively, we could instead choose to set $ 2 a e^{ - 2 \Phi_{0} } $
equal to a
constant (a freedom
we would seem to have if we can set $a = 1 $), in which case
the thermodynamic mass is simply proportional to $M$, which we have modelled
with a point source of dust and of dilaton at the origin \rq{29}--\rq{31a}.
In other words, our collapse solution motivates us to identify $M$ itself as
the thermodynamic mass of the black hole. In this context we treat $M$ not as
a fixed parameter, but as a variable strength for the localized
source which is breaking
down the conformal invariance. This mass identification
is analogous to that of ref.\ \cite{Arnold} and also agrees with
the mass one gets by taking the Newtonian limit of this theory. Making this
identification of mass, and integrating \rq{63}, we get
\be
S_{bh} = \frac{2\pi}{\hbar}\ln(\frac{M}{M_0}), \label{eq64}
\ee
where $M_0$ is a constant of integration, yielding
a logarithmic dependence of black-hole entropy on mass.
Note that this result is
identical to that in \cite{MST}. The constant $M_0$ appears as a fundamental
length in the two-dimensional theory, similar to the Planck mass in the
four-dimensional case. More details on the entropy of black holes in two
dimensions
are given in \cite{MST}.

As a first step in determining more fully the quantum properties of
the black hole, we turn now to another viewpoint, the 't Hooft ``brick wall''
treatment of the horizons. To avoid difficulties arising at the horizon,
't Hooft has suggested \cite{hooft} that particle wave functions near a
(four-dimensional) horizon are modified by gravitational interactions between
the incoming and outgoing waves, causing an effective cutoff of wavemodes near
the horizon. Modelling the suggested interactions with a simple ``brick wall''
cutoff a short distance from the horizon, he matched the normal Hawking
properties of the black hole and showed that the cutoff distance was an
intrinsic property of the horizon, independent of the mass of the source. In
\cite{MST}, this cutoff was examined for the simple black hole (\ref{eq16}),
for which
the cutoff distance is found to depend on both the mass of the source and the
rest mass of the particle.

We now examine a two-dimensional ``brick wall'' cutoff for \rq{32}.
Assume that the wave functions vanish within some constant distance $h$ of the
horizon,
\be
\phi(x)=0 \mbox{ if }x\leq x_H+h,    \label{eq65}
\ee
where $\phi(x)$ is the scalar wave function of a particle mass $m$. Now take
$\phi$ to be a stationary state, in which case the Klein-Gordon equation
implies
that
\be
(\alpha\phi')' + [E^2\alpha^{-1} - m^2]\phi  = 0. \label{eq66}
\ee
Changing to coordinates such that $dz=\alpha^{-1}dx$, with an overdot denoting
$d/dz$, we get
\be
\ddot{\phi}+[E^2\alpha^{-1}-m^2]\phi =0. \label{eq67}
\ee
Using a WKB approximation $\phi=e^{iS}\varphi$, we find that the radial
wavenumber $k(x,E)=dS/dx$ is given by
\be
k^2=\frac{1}{\alpha^2}[E^2-m^2\alpha],   \label{eq68}
\ee
which is valid as long as the right-hand side is non-negative. The density of
states is given by the total number of radial modes $\pi N =\int dx\, k(x,E)$,
\be
g(E) = \int\frac{dx}{\alpha(x)}\sqrt{E^2-m^2\alpha(x)}, \label{eq69}
\ee
integrating from the brick wall $x=x_H+h$ up to the maximum given by
$E^2-m^2\alpha(x)=0$. This gives
\begin{eqnarray}
g(E)&=&\frac{1}{2M}\left[ E\ln\left(
\frac{2E^2+\gamma+2E\sqrt{E^2-\gamma}}{\gamma}
\right)\right.\nonumber\\
 & &-\left.\sqrt{E^2-m^2}\ln\left(
\frac{2E^2-m^2-\gamma+2\sqrt{E^2-m^2}\sqrt{E^2-\gamma}}{m^2-\gamma}
\right)\right]    \label{eq70}
\end{eqnarray}
where $\gamma = m^2(1-e^{-2Mh})$. Approximating for small $m$ ($m\ll E$) yields
\be
g(E) = -\frac{E}{2M}\ln(e^{2Mh}-1).  \label{eq71}
\ee


As the particle is a boson, the free energy $F$ at an inverse
temperature $\beta$ is given by
\be
e^{-\beta F} = \prod_{N} \frac{1}{(1-e^{-\beta E_N})},  \label{eq72}
\ee
yielding
\begin{eqnarray}
\pi\beta F& = &\int dg(E) \ln(1-e^{-\beta E})       \nonumber      \\
 & = &-\beta\int_{E_0}^{\infty}\frac{e^{-\beta E}}{1-e^{-\beta E}}g dE,
\label{eq73}
\end{eqnarray}
where $E_0=m\sqrt{1-e^{-2Mh}}$ is the energy necessary
to reach the brick wall. Substituting (\ref{eq71}) for $g$, and neglecting
$E_0$ (of order $m$), we get
\be
F=\frac{1}{2\pi M}\int_{0}^{\infty} \frac{EdE}{e^{\beta E}-1} \ln(e^{2Mh}-1).
\label{eq74}
\ee
The solution is
\be
F=\frac{\pi}{12M\beta^2}\ln(e^{2Mh}-1), \label{eq75}
\ee
where we have neglected all terms of linear and higher order in $m$. The free
energy in
(\ref{eq75}) may be identified with the free energy of the black hole itself.
Note that the free energy is independent of $m$, unlike the free energy of the
black hole in \cite{MST}. The total energy and entropy are
\begin{eqnarray}
U&=&\frac{\partial}{\partial\beta}(\beta F)= -\frac{\pi}{12M\beta^2}
\ln(e^{2Mh}-1),  \label{eq76}\\
S&=&\beta(U-F)= -\frac{\pi}{6M\beta}\ln(e^{2Mh}-1).  \label{eq77}
\end{eqnarray}
Note these are both positive, as $2Mh\ll 1$.

We now choose the cutoff distance $h$ so that the entropy $S$ and inverse
temperature $\beta$ are given by (\ref{eq63a} or \ref{eq64}) and (\ref{eq61})
respectively, so that we can identify them with the Hawking values
\cite{hooft}.
We can try this first by equating (\ref{eq63a}) with (\ref{eq77}) and
$\beta=2\pi/M$. Then we obtain
\be
\frac{1}{12}\ln(e^{2Mh}-1)= \frac{2\pi a}{M}
\label{eq77a}
\ee
which gives
\be
h=\frac{1}{2M}\ln\left(1+e^{- \frac{24 \pi a}{M} } \right). \label{eq77b}
\ee

The invariant distance from the horizon to the brick wall is
\begin{eqnarray}
ds &=&
\int_{x_H}^{x_H+h}\frac{dx}{\sqrt{\alpha}} \nonumber\\ &=& -
\frac{1}{2M}\ln\left[ 1-2\left(
e^{ \frac{-24\pi a}{M}}      \!\!\!+
e^{ \frac{-48\pi a}{M}}    \right)^{1/2}      \!\!\! +
2e^{\frac{-24\pi a}{M} } \right]\!
. \label{eq77c}
\end{eqnarray}
We thus obtain the result that the invariant cutoff distance depends upon
the strength of the source (although not on the mass of the particle) and
is not an intrinsic feature of the horizon. The total energy $U$ is
\be
U=\half a \! . \label{eq77d}
\ee
If we instead equate (\ref{eq64}) with (\ref{eq77}), we have
\be
-\frac{1}{12}\ln(e^{2Mh}-1)=2\pi\ln\left(\frac{M}{M_0}\right),
\label{eq78}
\ee
which gives
\be
h=\frac{1}{2M}\ln\left(1+\left(\frac{M}{M_0}\right)^{-
24\pi}\right).\label{eq79}
\ee
The invariant distance from the horizon to the brick wall is then
\begin{eqnarray}
ds &=&
\int_{x_H}^{x_H+h}\frac{dx}{\sqrt{\alpha}} \nonumber\\ &=& -
\frac{1}{2M}\ln\left[ 1-2\left(\left(\frac{M}{M_0}\right)^{-24\pi}\!\!\!+
\left(\frac{M}{M_0}\right)^{-48\pi}\right)^{1/2}\!\!\! +
2\left(\frac{M}{M_0}\right)^{-24\pi}\right]\!. \label{eq80}
\end{eqnarray}
Again, the invariant cutoff distance depends upon
the strength of the source and
is not an intrinsic feature of the horizon. The total energy $U$ is now
\be
U=\half M \ln\left(\frac{M}{M_0}\right)  \label{eq81}
\ee
which is identical to the result in \cite{MST}.

\section{Inclusion of Tachyons}

In the sigma-model approach to the string theory, consideration of the $L_0$
operator acting on the tachyon gives the following metric \cite{VerDijk}:
\be
ds^2 = \half(k-2) [dr^2-4 (\coth^2\frac{r}{2}-\frac{2}{k})^{-1}d\tau^2].
\label{veq3}
\ee
As $\alpha' = k-2$, we have $Q^2/4 = 2/(k-2)$. This solution, unlike the
solution (\ref{eq1}) of the simpler theory, is naturally symmetric. We would
like to write this metric in the same form as (\ref{eq1}). We make the
coordinate transformation
\be
t = \frac{2}{Q} \tau,   \label{veq4}
\ee
\be
x = -\frac{2}{Q} \ln\left(\sqrt{\cosh^2(r/2)+Q^2/4} + \cosh(r/2)\right),
\label{veq5}
\ee
which puts the metric in the form
\be
ds^2 = 4(-\alpha dt^2 + \alpha^{-1}dx^2),
\alpha(x) = 1-\frac{k}{1+\cosh(Qx-\ln(2k-4))}. \label{veq6}
\ee
We can absorb the overall factor of 4 in this metric by a redefinition of the
proper length. We then see that this reduces to (\ref{eq1}) as $k\to\infty$,
since the constant $a$ in \rq1 can be suppressed by translating $x$.
For $k$ finite, a further translation of $x$ will put the new metric
(\ref{veq6}) in the form
\be
ds^2 = -\alpha dt^2 + \alpha^{-1} dx^2,
\alpha(x) = 1-\frac{k}{1+\cosh(Q x)}. \label{veq7}
\ee
We see that this form of the black hole is symmetric and asymptotically flat as
$x\to\pm\infty$. The horizons are at $|x|=x_H = \cosh^{-1}(k-1)/Q$
(note $x_H\to\infty$ as $k\to \infty$). We will now briefly examine the
properties of this black hole.

We first consider the motion of point particles. As the form of the metric
is the same, the equations of motion will be (\ref{eq43},\ref{eq44}),
which give
\be
\dot{t}=1/\alpha,\,\, \dot{x}^2 = 1-E^2\alpha,   \label{veq8}
\ee
where for massless particles $E=0$, for massive particles $E^2>0$. We will
restrict to $|x|>x_H$ while solving these equations. We proceed to
solve the equations for massless particles, yielding the explicit solution
\begin{eqnarray}
|x|(\lambda) &=& -\lambda + x_0 \label{veq9} \\
t(\lambda) &=& \lambda
+ \sqrt{\frac{k}{2}} \tanh^{-1}\left(\sqrt{\frac{k}{k-2}}
\tanh\left(-\frac{Q}{2} (\lambda-x_0)\right)\right) + C, \label{veq10}
\end{eqnarray}
where $C$ is a constant of integration. As $x\to x_H$, $\cosh(Qx)\to k-1$, and
thus the coordinate time diverges, $t\to\infty$, as expected.

For massive particles, the solution of the equation for $x$ is again divided
into three cases. For simplicity, we will
give explicit solutions for $x$ in terms of the proper time $\tau$ for each
case, and a general form for the coordinate time $t$ in terms $\tau$ and $x$.
For the case $E^2<1$, the solution for $x$ is
\be
|x|(\tau) = \frac{1}{Q}\cosh^{-1}\left(
\frac{(E^2k+2-2E^2)\cosh(E\sqrt{1-E^2}Q \tau)-E^2k}{2-2E^2}\right)+x_0.
\label{veq11}
\ee
For the case $E^2=1$,
\be
|x|(\tau) = \frac{1}{Q} \cosh^{-1}( \frac{Q^2k}{4} \tau^2 +1)+x_0.
\label{veq12}
\ee
Finally, for $E^2>1$, where we have to restrict to $x$ such that
$1-E^2\alpha(x) > 0$, the solution is
\be
|x|(\tau) = \frac{1}{Q} \cosh^{-1}\left(
\frac{(E^2k+2-2E^2)\cos(E\sqrt{E^2-1}Q \tau) -E^2k}{2-2E^2}\right)+x_0.
\label{veq13}
\ee
The solution for the coordinate time is expressed in terms of the proper time
$\tau$ and the function $\xi(x) = (\cosh(Qx)+1-k)^{-1}$. The solution in all
cases is
\be
t = \tau + \frac{k}{EQ\sqrt{k(k-2)}}\ln(2\sqrt{a}\sqrt{(1-E^2)+b\xi+a\xi^2}
+2a\xi +b)+C, \label{veq14}
\ee
where $a=k(k-2), b=-E^2k+2(k+E^2-1)$ and $C$ is a constant of integration.
As $|x|\to x_H$, $\xi\to\infty$, and thus
the coordinate time $t\to\infty$, as expected. Thus the qualitative form of
point particle motion is consistent with our previous results. As before, these
solutions only apply to initially infalling particles outside the black hole.

We  next consider the quantum properties of this system. The
temperature of the horizons may be calculated as before \cite{BHT}, and we find
\be
T= \frac{\hbar}{2\pi}\left|\frac{\alpha'(x_H)}{2}\right| =
\frac{\hbar}{2\pi} \sqrt{2/k}. \label{veq15}
\ee
The consistent temperature results from the two other black holes previously
considered prompt us to identify the mass from this temperature as
$M =\sqrt{2/k}$. This reduces to
the value $M=Q/2$ given for the simpler string theory black hole
(\ref{eq32}) in the limit $k\to\infty$.
If we now consider the entropy with
this mass identification, and write only the analog to (\ref{eq64}), we obtain
\be
S_{bh} = -\frac{\pi}{\hbar}\ln\left(\frac{k}{k_0}\right),
\label{veq16}
\ee
where $k_0$ is a constant of integration having
dimensions of an inverse length squared. Turning to the 't Hooft brick wall
model \cite{hooft}, we see that the integral for the density of states will
have the form (\ref{eq69}), which may be rewritten as
\be
g(E) =
\frac{\sqrt{k}}{Q} \int d\alpha \frac{\sqrt{E^2 -m^2\alpha}} {\alpha(1-
\alpha) \sqrt{k-2+2\alpha}}.  \label{veq17}
\ee
If we integrate between the
limits $\alpha = E^2/m^2$ and $x=x_H+h$ and neglect all terms in $m$, we
find that the solution is
\begin{eqnarray}
g(E)& = &-\frac{E}{Q} \left(\ln(\sqrt{\gamma^2-2\gamma}+\gamma-1) \right.
+\sqrt{\frac{k}{k-2}} \ln(1-\gamma/k)  \label{veq18} \\
& & -\sqrt{\frac{k}{k-2}} \ln(\sqrt{1-2/k}\sqrt{\gamma^2-2\gamma} \left.
+\gamma(1-\frac{1}{k})-1)  \right),
\nonumber
\end{eqnarray}
where $\gamma = 1+\cosh(Qx_H+Qh)$. If we now
substitute this form of $g$ in the integral (\ref{eq73}) for the free
energy $F$, we find that
\be
F = \frac{\pi}{6Q\beta^2} f(h,k),
\label{veq19}
\ee
where $f(h,k) = -Qg(E)/E$. The entropy is then
\be
S = -\frac{\pi}{3Q\beta} f(h,k).    \label{veq20}
\ee
Comparing this
entropy to (\ref{veq16}), using $\beta=T^{-1}$, will give $h$ in terms of
$k$. This is given implicitly by the equation
\be
f(h,k) = 12\pi \sqrt{1-\frac{2}{k}}\ln\left(\frac{k}{k_0}\right). \label{veq21}
\ee
So the value of $h$, and thus the invariant distance, will depend on $k$ in
a highly nonlinear fashion.

\section{Conclusions}

Locally the black hole solution \rq1 proposed in \cite{JurgJMP,MSW} has
(for $x>0$) many qualitative and quantitative similarities to the simple
black hole \rq{16} \cite{MST,Arnold}.  We have seen that in order to
understand this solution as being the exterior metric outside a
symmetric distribution of matter, it is necessary to modify it by
replacing $x$ with the two dimensional
``radial'' variable $|x|$, something quite natural in the context of a
two-dimensional theory of gravity based on the action (\ref{eq11}).
The resulting black hole (\ref{eq32}) has the same local features as \rq1,
and is similar in many respects to the black hole \rq{16} studied
previously \cite{MST,Arnold}.  However, while the exterior metric \rq{32} is
asymptotically flat, the metric \rq{16} is (locally) exactly flat, as it
corresponds to two Rindler spaces of opposite acceleration parameter
matched to a symmetric distribution of matter \cite{MFound}.

The modification from $x$ to $|x|$ significantly changes the properties of
the spacetime investigated in \cite{MSW,EDW}.   While the original
asymmetric black hole \rq1 is identical to the black hole \rq{32} in the
region outside its horizon (for positive $x$),  the identification of the
``mass'' of the black hole in the former case is made in the context of the
massive scalar field rather than implying a strictly physical source. The
exponential singularity in the curvature at $x\to\infty$ becomes a delta-
function singularity at the origin. A localized physical source such as
gravitationally collapsing dust can give rise to this black hole only if
appropriate surface stress energies and dilaton surface charges are
included.  The latter may be understood to arise from a tachyon potential
at the edges of the dust.  Hence we see that inclusion of tachyonic effects
(albeit in a phenomenologically simple form) is necessary to ``symmetrize''
the string black hole.

The more realistic inclusion of tachyon effects resulting in the more
complicated black hole solution (\ref{veq7}) proposed in \cite{VerDijk} is
shown (after a translation) to be symmetric about the origin. It includes
the simpler black hole \rq{32} in the $k\to\infty$ limit. Their properties
are similar in that each has the expected properties for point particle
motion, and if we identify the mass as $M=\sqrt{2/k}$ the quantum
properties are also the same.

The similarities between the black hole solutions investigated here and the
$(3+1)$ dimensional Schwarzchild case suggest that the theory \rq{11} coupled
to matter should have other interesting properties. Work on this area is in
progress.

\section{Acknowledgments}

This work was supported in part by the Natural Sciences and Engineering
Research Council of Canada. Dr. M. Morris wishes to thank the Canadian
Institute for Theoretical Astrophysics for partial support. We would
especially like to thank W. Israel for discussions concerning the matching
conditions. The integrals in this paper were evaluated using the Maple 4.3
symbolic algebra package.


\begin{thebibliography}{References}
\bibitem{BHT}J.D. Brown, M. Henneaux and C. Teitelboim, Phys. Rev. {\bf D9}
(1974) 3292; Lett. Nuovo Cimento {\bf4} (1972) 737
\bibitem{MST}R.B. Mann, A. Shiekh, and L. Tarasov, Nucl. Phys. {\bf B341}
(1990) 134.
\bibitem{MFound}R.B. Mann, Found. Phys. Lett. (to be published).
\bibitem{Arnold}A.E. Sikkema and R.B. Mann, Class. Quantum Grav. {\bf 8}
(1991) 219.
\bibitem{semi}R.B. Mann, S.M. Morsink, A.E. Sikkema and T.G. Steele,
Phys. Rev. {\bf D43} (1991) 3948.
\bibitem{SharTomRobb}S.M. Morsink and R.B. Mann, Class. Quant. Grav.
(to be published); R.B. Mann and T.G. Steele, Class. Quant. Grav.
(to be published).
\bibitem{JurgJMP} H.-J. Schmidt, J. Math. Phys. {\bf 32} (1991) 1562
\bibitem{L+R}A.R. Lugo and J.G. Russo, Stanford University preprint, SU-ITP-896
\bibitem{MSW}G. Mandal, A.M. Sengupta, and S.R. Wadia, IAS preprint,
IASSNS-HEP-91/10.
\bibitem{EDW}E. Witten, IAS preprint, IASSNS-HEP-91/12
\bibitem{napyost}M. D. McGuigan, C. R. Nappi, S. A. Yost, IAS preprint,
IASSNS-HEP-91/57.
\bibitem{BanksL}T. Banks M. O'Loughlin, Nucl. Phys. {\bf B362} 649 (1991).
\bibitem{VerDijk} R. Dijkgraaf, H. Verlinde and E. Verlinde, Princeton preprint
PUPT-1252, IAS preprint IASSNS-HEP-91/22.
\bibitem{entropy}J.D. Bekenstein, Phys. Rev. {\bf D7} (1973) 2333
\bibitem{hooft}G. 't Hooft, Nucl. Phys. {\bf B256} (1985) 727
\end{thebibliography}
\end{document}